\documentclass[12pt]{article}

\usepackage[dvips]{graphicx}

\usepackage{SIunits}

\begin{document}

\title{Numerical optimization of a grating coupler for the efficient excitation of surface plasmons at an Ag-SiO$_2$ interface}

\author{Jesse Lu\footnote{email:jesselu@stanford.edu}, Csaba Petre, Eli Yablonovitch\\UCLA Electrical Engineering Dept.,\\Los Angeles CA 90095-1594\\\\Josh Conway\\The Aerospace Corporation\\El Segundo, CA 90245-4691}

\maketitle

\begin{abstract}
The efficient generation of surface plasmons from free space optical waves is still an open problem in the field. Here we present a methodology and optimized design for a grating coupler. The photo-excitation of surface plasmons at an Ag-SiO$_2$ interface is numerically demonstrated to yield greater than 50\% coupling from a Gaussian beam into surface plasmon voltages and currents.
\end{abstract}

\section{Introduction}
Surface plasmons are optical-frequency, alternating-currents that propagate along the surface of a conductor~\cite{Raether1988}. These optical frequency surface modes enable metallic structures to focus light to the nanoscale~\cite{Conway2006}, enabling an enhancement and confinement of the electric field far beyond what is possible in dielectric optics. The tremendous potential for energy concentration and localization spans a multitude of disciplines and has the potential to enable a host of future technologies. For instance, surface plasmon-based instrumentation is now commonplace in chemical spectroscopy and in detection of trace biological material through surface-enhanced Raman scattering~\cite{Holoma2003}, where Raman enhancements as high as $10^{14}$ have been reported.~\cite{Nie1997}~\cite{Kneipp1997} These astronomical enhancements occur at stochastically located `hot spots' or on resonances of engineered nanoparticles~\cite{Su2006}. The prime mechanism behind this effect is the coupling of free-space optical energy to a localized plasmonic standing wave. Exploiting these large field enhancements for many engineering applications, however, requires coupling to a plasmonic traveling wave. Applications include Heat-Assisted Magnetic Recording for the next-generation of hard disks~\cite{Challener2003}, near-field lithography~\cite{Luo2004}, surface plasmon microscopy~\cite{Rothenhausler1988}, and sub-wavelength optical circuitry~\cite{Bozhevolnyi2006}. In such plasmonic devices the efficient excitation of surface plasmons is critically important in order to ensure a strong electric field enhancement. Indeed, inefficient coupling may even nullify the electric field enhancement which would otherwise be obtained by such devices and this critical coupling is still considered an open problem in nano-optics.
\\
Schemes for the conversion of free-space optical waves into currents and voltages include coupling by attenuated total reflection in prisms, end-fire coupling to plasmonic waveguides~\cite{Stegeman1983}~\cite{Charbonneau2005} and various antenna structures.\cite{Cubukcu2006}~\cite{Siu1977} However, exciting plasmons by attenuated total reflection requires a dielectric prism which is a macroscopic device and therefore too bulky for most applications. Efficient end-fire coupling, on the other hand, is limited in applicability to special waveguide structures which are not compatible with planar fabrication technologies. Most optical antenna structures suffer from low radiation resistance which leads to Ohmic losses and a low directivity radiation pattern which provides a poor mode match to incident plane waves. The grating coupler then emerges as a most viable method to efficiently convert laser power to electrical voltages and currents. Indeed the plasmon grating coupler can be considered a special case of antenna, analogous to a periodic antenna array or Yagi-Uda antenna~\cite{Stutzman1998} which achieve high efficiencies due to very directive beam formation.
\\
The high directivity associated with large grating antenna structures leads to a high antenna capture cross-section, given by $\sigma_{antenna}=\lambda^2 / \Omega$, where $\Omega$ is the acceptance solid-angle, and $\lambda$ is the vacuum wavelength.  Thus the captured power is $P_{in} = \textrm{Intensity} \cdot ( \lambda^2 / \Omega ) $, which can be converted to an optical frequency voltage by the formula $V^2 = P_{in} \cdot R_{rad}$, where $R_{rad}$ is the antenna radiation resistance which is typically less than $50$   Ohms.
\\
Grating couplers have been employed in previous experimental work and some rough optimizations have been implemented.\cite{Peng2004}~\cite{Leveque2005} However, a thorough effort to optimize a grating coupler has not yet been presented. In this paper, we present the methodology and design of an optimized grating coupler topology which couples 50\% of the energy from a 476 nm free-space wavelength incident optical beam into the surface plasmon mode at an Ag-SiO$_2$ interface.

\section{Mathematical Framework}
\begin{figure}
\centering
\includegraphics{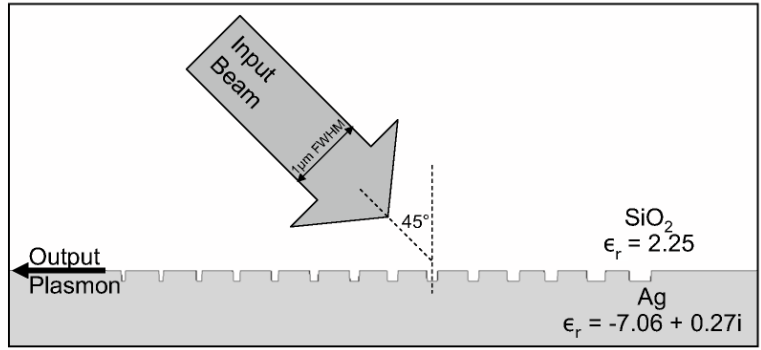}
\caption[Setup of the optimization problem]{An input beam with a Gaussian profile strikes the grating at 45 degrees off incidence and is coupled contra-directionally into the surface plasmon mode.\label{Setup}}
\end{figure}

\begin{figure}
\centering
\includegraphics{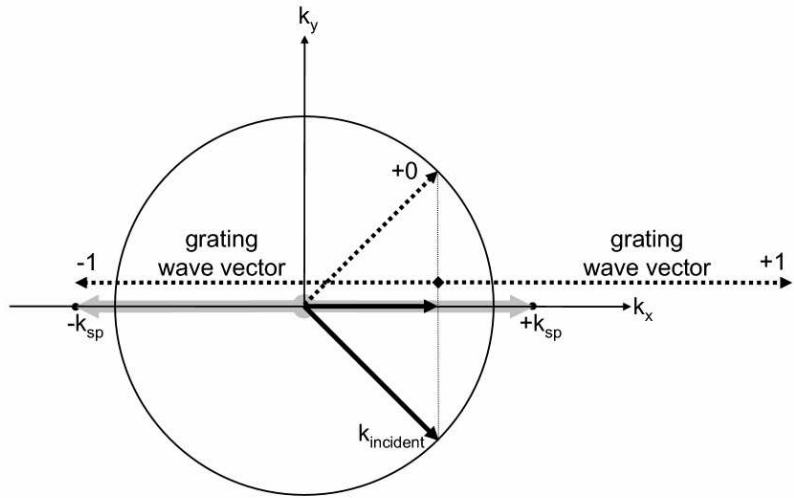}
\caption[Wave-vector matching to the contra-directional surface plasmon mode.]{Wave-vector diagram illustrating how a $(-1)$ grating wave vector is added to the parrallel component of the wave-vector of the incident beam in order to match the wave-vector of the contra-directionally propagating surface plasmon mode $(-k_{sp})$. The advantage of coupling contra-directionally is that only the specular reflection order $(+0)$ needs to be suppressed.\label{Kdiagram}} 
\end{figure}

The design space of the grating was limited to a one-dimensional array of rectangular grooves of uniform depth at an Ag-SiO$_2$ interface (Figure~\ref{Setup}).  The dielectric constants corresponding to a free-space wavelength of 476 nm were $\epsilon_r = -7.06 + i0.27$ and $\epsilon_r = 2.25$, for Ag and SiO$_2$ respectively~\cite{Conway2006}.  The imaginary part of the permittivity of Ag was included in order to take into account resistive losses in silver.  The input beam was chosen to be of a Gaussian profile with a FWHM diameter of \unit{1}{\micro\meter} and incident on the grating at a 45 degree angle.  Additionally, we chose to couple contra-directionally to the surface plasmon mode in order to eliminate all diffraction orders, leaving only the specular reflection to be suppressed, allowing all the remaining power to be transferred into optical frequency surface currents (Figure~\ref{Kdiagram}).
\\
Numerical simulations were obtained using COMSOL, a finite-element-method solver, concentrating on the transverse magnetic (TM) mode. Perfectly matched layers, where the imaginary parts of the permittivity and permeability increase as a cubic power, were used as absorbing boundaries~\cite{Conway2006}.  Computational instabilities associated with the sharp corners on the grating were avoided by convolving a “square-tooth” topology with a Gaussian function of standard deviation 1 nm.
\\
We defined the efficiency of the grating coupler as the total power of the coupled surface plasmon as it exited the grating divided by the total power of the input beam.  The power of the coupled surface plasmon was first computed several microns away from the nearest edge of the grating by integrating the value of the Poynting vector along a line perpendicular to the Ag-SiO$_2$ interface; the power in the plasmon wave at the grating edge was then calculated using the known decay length of the surface plasmon.  It was specifically this coupled plasmon power calculated at the edge of the grating which was employed in the efficiency calculation.  The power in the input beam was computed by simply integrating for the Poynting vector over the cross section of the beam. For this input power calculation the silver was replaced with appropriate absorbing boundaries in order to eliminate backscattering.
\section{Problem Challenge}
From a mathematical standpoint, converging on an optimal grating topology is non-trivial. This is because of both the size and complexity of the optimization space. 
\\
The expansiveness of the optimization space can be grasped when one considers that a reasonably efficient topology must be the result of an 18-parameter optimization, at the very least.  One arrives at this minimum number of parameters by first considering that the cross-section of the beam requires about nine grooves for correct wave vector matching~\cite{Raether1988}.   This corresponded to a daunting 18-parameter optimization since each groove must be characterized by at least its width and spacing.  This overly large optimization space rendered genetic optimization algorithms and other direct search algorithms~\cite{Lewis2000} completely ineffective in finding a meaningful optimum.
\\
In addition to the expansiveness of the optimization space, the unavoidable presence of multipath interference in electromagnetic wave problems further exacerbated the task of mathematically searching for an optimum.  Owing to the multipath interference, we found that the optimization space was studded with numerous local optima. The situation was analogous to that of laser speckle, where although there is a global maximum at the center where the laser light is most intense, there also exist numerous local maxima known as speckle surrounding the center. It is therefore impossible for traditional numerical methods such as steepest descent or simplex optimization methods, which converge on local optima, to arrive at the global optimum unless by a fortuitous starting point in the immediate vicinity of the global optimum.
\\
The conclusion of these two observations is that one can arrive at an optimal topology only if one has an initial topology which is already quite close to the optimum.  The crux of the problem is therefore in the successful selection of such an already-efficient initial topology.

\section{Methodology}
To determine a near-optimal initial topology, a hierarchal optimization method was used which was repeated over a range of beam input angles, beam input widths and groove depths. First, we started with a grating consisting of a single groove with strong coupling characteristics placed at the edge of the space where the grating was to be constructed, opposite the edge from which the surface plasmon would exit the grating.  Subsequent grooves were then placed roughly one grating period away by performing a two-parameter search which determined the most efficient width and position for such a groove, under the condition that previously placed grooves remained fixed.  Additional grooves were added sequentially in this way until the efficiency of the grating could no longer be improved. Note that the grating design was completely deterministic after the initial placement of the first groove, since the successive two-parameter groove optimization that was used was deterministic as well.
\\
Having produced a reasonably efficient design, the width and position of each groove was adjusted individually, starting from the rear-most groove and then moving forward, using the same two-parameter search as in the initial placement phase.  Lastly, the grating was fine-tuned using a generic direct search algorithm in conjunction with numerical simulation results.
\\
The two-parameter search for each successive groove consisted of calculating the efficiency of the grating while varying the location and width of each groove over some range of values and then selecting the most efficient configuration.  However, individually determining the efficiency of every configuration by numerical simulation was overly tedious computationally.  To make the search many times faster, we modeled the grating as a transmission line studded with scattering centers~\cite{Narasimha2004}. This was accomplished by using numerical simulation to tabulate the scattering parameters of individual grooves of varied widths and then computing the strength of the output plasmon wave by using a simple transfer matrix method~\cite{Helszajn1992}.  

\section{Results}
\begin{figure}
\centering
\includegraphics{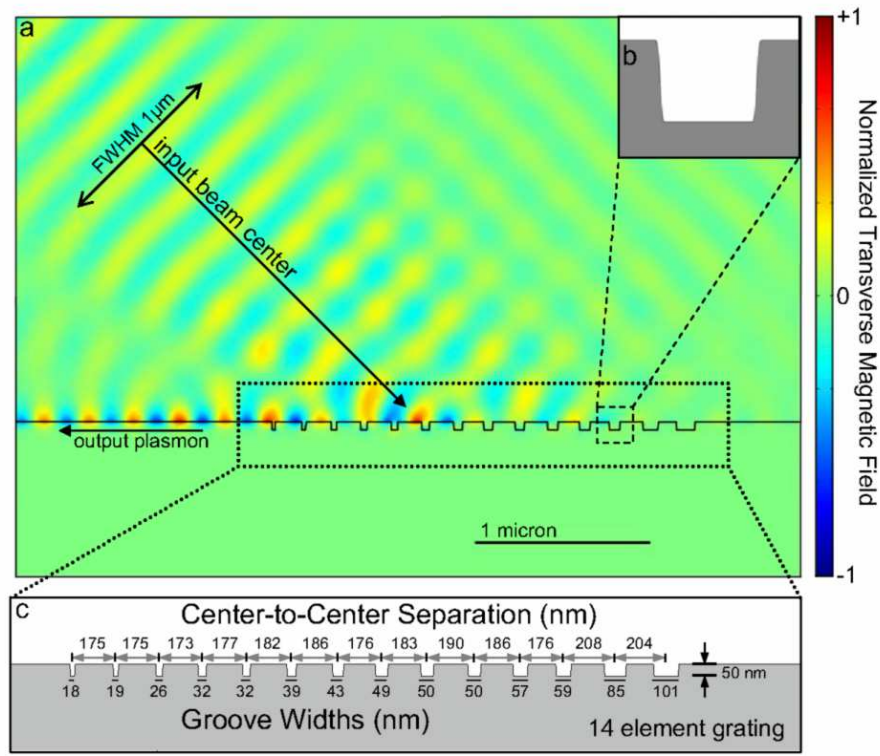}
\caption[The optimized grating.]{The optimized grating. \textbf{a)} Finite Element Method simulation of the TM magnetic field. \textbf{b)} Close-up of a groove showing the slight rounding of the corners needed to stabilize the FEM simulation. \textbf{c)} Diagram of the widths and center-to-center separation distances of the grooves which form the grating.\label{Optimized Grating}}
\end{figure}

\begin{figure}
\centering
\includegraphics{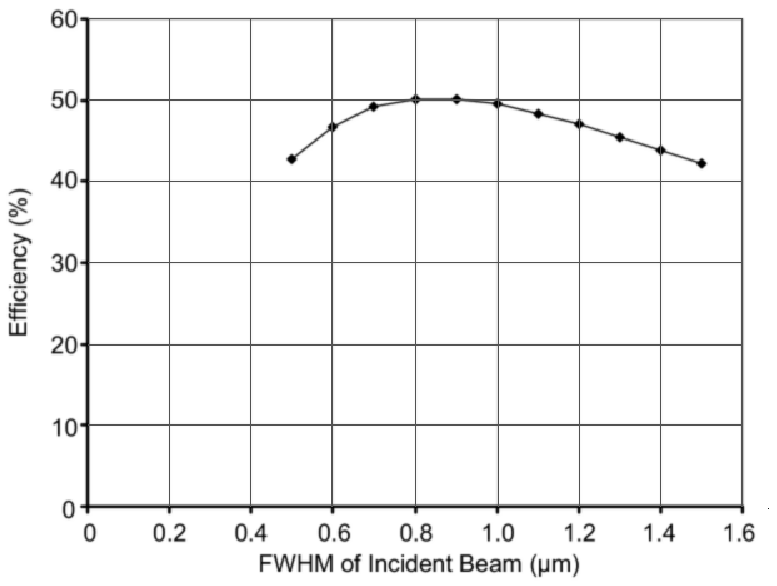}
\caption[Robustness to variations in input beam width]{The diameter of the input beam can be varied substantially without significantly decreasing the efficiency of the grating.\label{Beamwidth Robustness}}
\end{figure}

\begin{figure}
\centering
\includegraphics{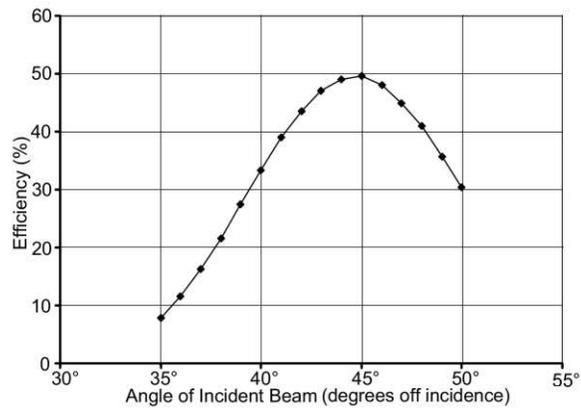}
\caption[Robustness to variations in input beam angle]{The efficiency of the optimized grating decreases rapidly with even small variations in the angle of the incident beam. This is expected due to the high directivity of the grating coupler acting as an antenna array.\label{Angle Robustness}}
\end{figure}

The final grating coupled 50\% of the energy from the input beam into the surface plasmon mode (Figure~\ref{Optimized Grating}).  Its final optimized structure consisted of a 14-element grating with a groove depth of 50 nm. The average center-to-center spacing between grooves was 184 nm, slightly larger than the expected value of 165 nm dictated by wave-vector matching~\cite{Raether1988}.  
\\
Also, the groove width of the optimized grating decreased monotonically from 101 nm to 18 nm from the rear to the front of the grating, which was precisely dictated by the need to match the Gaussian profile of the input beam.  To elucidate this further we consider the time-reverse case where the plasmon propagates to the right, strikes the front of the grating and then is coupled to free space radiation. In this case, one would expect the front of the grating to be composed of weak scattering centers and the middle of the grating to be composed of stronger scattering centers in order to produce a beam with a Gaussian profile, since such a beam's energy is more concentrated toward its center.  However, toward the rear of the grating, even stronger scattering centers are needed, since the strength of the incoming surface plasmon has also decreased substantially as well. Thus the rearmost grooves, in order to produce the desired Gaussian profile, must do their best to scatter all the remaining surface plasmon mode energy.
\\
In order to validate our optimized grating, its robustness to small topological changes was calculated.  To this end we found that the width of any individual groove could be varied by up to 5 nm without incurring the loss of more than 5\% efficiency.  The location of any individual groove could be shifted by 10 nm and the depth of the grating could also be varied by 5 nm without losing more than 5\% efficiency as well. Also, we found that the use of a Gaussian apodization function with standard deviation either twice or half that of the original would not result in a loss of more than 5\% efficiency either. These robustness calculations confirmed that our topology was indeed a valid, physical optimum, and not a mere computational artifact.
\\
The sensitivity of the grating to changes in the width, alignment and angle of the incoming beam was also calculated.  Figure~\ref{Beamwidth Robustness} shows that the grating was relatively insensitive to the diameter of the beam in that the grating remained more than 45\% efficient whether the beam width was narrowed to \unit{0.6}{\micro\meter} or widened to \unit{1.3}{\micro\meter}. In addition to being insensitive to the width of the beam, we found that the grating was also insensitive to the alignment of the beam since a shift of up to 200 nm in the center position of the beam would not result in an efficiency below 45\%. Finally, because a grating is equivalent to a highly directive antenna array, we expected and found an extreme sensitivity to the incident beam angle.  Figure~\ref{Angle Robustness} shows that the beam angle could not deviate more than 2 degrees if we wished to maintain an efficiency greater than 45\%. 
\section{Conclusion}
We have presented a grating coupler topology optimization that couples 50\% of the energy from a \unit{1}{\micro\meter} diameter beam (FWHM) at $\lambda = 476$ nm free space wavelength, into surface currents and voltages at an Ag-SiO$_2$ interface.  The angle of incidence was chosen to be 45 degrees, and chosen to couple into the surface plasmon mode contra-directionally.  We note the difficulty caused by multiple optimization parameters as well as multipath in electromagnetic optimization, and we present an optimization routine that overcomes these problems.  Also, we show that our design approach is robust to topological variations as well as to significant changes in the beam diameter and alignment.  We verify that the grating is highly directive, as expected.
\section{Acknowledgment}
A portion of this work was peformed as part of The Aerospace Corporation's Independent Research and Development Program.

\end{document}